\documentclass[5p,times,10pt,nofootinbib,twocolumn]{revtex4-2}

\usepackage{graphicx,natbib,lineno,color}
\usepackage{amsmath}

\begin{document}

	\title{Emergence of quadratic gravity from entanglement equilibrium}
	
	\author{Ana Alonso-Serrano}
	\email{ana.alonso.serrano@aei.mpg.de}
	\address{Max-Planck-Institut f\"ur Gravitationsphysik (Albert-Einstein-Institut), \\Am M\"{u}hlenberg 1, 14476 Potsdam, Germany}
	
	\author{Marek Li\v{s}ka}
	\email{liska.mk@seznam.cz}
	\address{Institute of Theoretical Physics, Faculty of Mathematics and Physics, Charles University,
		V Hole\v{s}ovi\v{c}k\'{a}ch 2, 180 00 Prague 8, Czech Republic}

	\begin{abstract}
		In this work, we derive the linearised equations of quadratic gravity from entanglement equilibrium of local causal diamonds. Rather than starting from the Wald entropy prescription (which depends on the gravitational Lagrangian), we employ a model independent approach based on the logarithmic corrections to horizon entanglement entropy. In this way, we are able to show the emergence of linearised quadratic gravity from entanglement equilibrium without using any a priori knowledge about gravitational dynamics. If the logarithmic correction to entropy has a negative sign, as predicted by replica trick calculations of entanglement entropy, we find that the quadratic gravity correction terms have the sign necessary to avoid tachyonic instabilities of the theory.
	\end{abstract}

	\maketitle

	\section{Introduction}
	
	The thermodynamic equilibrium conditions applied to local, observer-dependent causal horizons encode gravitational dynamics~\cite{Jacobson:1995,Chirco:2010,Padmanabhan:2010,Svesko:2018}. This implied a further step in interconnecting thermodynamics and gravitational dynamics and opened the door for deeper studies of gravitational dynamics with well-defined thermodynamic tools. The key assumption for these analyses is that the local causal horizons posses Bekenstein entropy, \mbox{$S_{\text{B}}=\mathcal{A}/4l_{\text{P}}^2$}~\footnote{Throughout the paper, we set $c=k_{\text{B}}=1$ but we leave $G$ and $\hbar$ explicit to keep track of the gravitational and quantum effects (so that the Planck length reads $l_{\text{P}}=\sqrt{G\hbar}$). We work in four spacetime dimensions with the metric signature $\left(-+++\right)$, where Greek letters are used for spacetime indices and Latin ones for spatial indices. Other conventions follow~\cite{MTW}.}, with $\mathcal{A}$ being the area of a spacelike horizon's cross-section. In other words, black hole entropy is assumed to be associated with the presence of a horizon, so that any other causal horizon has entropy given by the same expression. Then, imposing the equilibrium condition $\delta S_{\text{B}}+\delta S_{\text{m}}=0$, where $\delta S_{\text{m}}$ denotes Clausius entropy of the matter crossing the horizon~\cite{Baccetti:2013}, implies the Einstein equations (for a discussion of details and the additional assumptions involved, see, e.g.~\cite{Chirco:2010,Alonso:2020a,Alonso:2021}). While this approach establishes a deep connection between gravitational dynamics and horizon thermodynamics, it has several drawbacks. These recently led to an improved approach based on the entanglement equilibrium of causal diamonds~\cite{Jacobson:2016}.
	
	One of the main concerns is the need to introduce Bekenstein entropy of a local causal horizon, whose microscopic origin is unclear. This is of course the case for black holes as well. However, the situation for local causal horizons is more complicated, since they are observer-dependent and the same must be true for their entropy (for arguments in favour of assigning entropy to local horizons, see e.g.~\cite{Jacobson:1994,Jacobson:2003,Baccetti:2013,Jacobson:2019}). This difficulty disappears, e.g. if one interprets black hole entropy in terms of quantum entanglement~\cite{Sorkin:1986,Srednicki:1993,Jacobson:2003,Solodukhin:2011}, as this description naturally assigns the same entropy to any causal horizon. Then, however, we employ quantum entanglement description for Bekenstein entropy, whereas entropy of the matter crossing the horizon is given by the (semi)classical Clausius formula. It is not obvious that the thermodynamic equilibrium condition can be correctly stated combining these two very different notions of entropy (although it has been shown that both entanglement and Clausius entropy of the matter flux coincide for causal diamonds~\cite{Carroll:2016,Alonso:2020a}).

	A more refined approach describes entropy of both geometry and matter in terms of quantum entanglement~\cite{Jacobson:2016}. Its starting point is the assumption that a vacuum, maximally symmetric spacetime is in entanglement equilibrium. The corresponding entanglement entropy is then extremal and its change due to a simultaneous variation of spacetime geometry and matter fields must vanish to the first order. In the presence of a local causal horizon, the entanglement entropy variation has two components~\cite{Jacobson:2016}. On the one side, the UV component, $\delta S_{\text{UV}}$, does not depend of the state of matter fields (by virtue of the strong equivalence principle) and, to the leading order, it is proportional to the variation of the horizon area, i.e., $\delta S_{\text{UV}}=\eta\delta\mathcal{A}$~\cite{Sorkin:1986,Srednicki:1993,Chirco:2010,Solodukhin:2011,Jacobson:2016}. The proportionality constant $\eta$ depends on the UV cut-off introduced to regularise the entropy and cannot be directly determined. On the other side, the IR contribution, $\delta S_{\text{IR}}$, depends on the state of the matter fields. If only conformally invariant matter fields are present, it turns out to be proportional to the energy-momentum tensor expectation value~\cite{Jacobson:2016,Jacobson:2019}. The entanglement equilibrium condition $\delta S_{\text{UV}}+\delta S_{\text{IR}}=0$ then leads to the Einstein equations, with the Newton gravitational constant defined in terms of the proportionality constant $\eta$, $G=1/\left(4\pi\hbar\eta\right)$. Then, we have $\eta=1/\left(4l_{\text{P}}^2\right)$ and the UV entanglement entropy must again correspond to Bekenstein entropy.
	
	Thermodynamic~\cite{Eling:2006,Padmanabhan:2010,Jacobson:2012,Svesko:2018} and entanglement equilibrium~\cite{Bueno:2017,Svesko:2019} derivations of equations for gravitational dynamics have been developed in the literature even for modified theories of gravity whose Lagrangian is an arbitrary function of the metric and the Riemann tensor. These approaches rely on the expressions for Wald entropy and generalised volume closely related with gravitational dynamics. Here, we explore a different direction to derive the modified equations for gravitational dynamics from entanglement equilibrium. Our aim is to avoid any reliance on the form of the gravitational Lagrangian. Thus, rather than starting from Wald entropy, we consider that the UV contribution to entanglement entropy contains a correction term logarithmic in the horizon area
	\begin{equation}
		\label{Se}
		S=\frac{\mathcal{A}}{4l_{\text{P}}^2}+\mathcal{C}\ln\frac{\mathcal{A}}{4l_P^2},
	\end{equation}
	where $\mathcal{C}$ is a real number whose value and sign are model dependent\footnote{For the case of round spheres and with only conformal matter fields present, $\mathcal{C}$ can be considered to be constant~\cite{Solodukhin:2011}. In the following, we only work with conformal matter and shape variations of the spherical horizon do not affect our analysis. Hence, we do not consider variations of $\mathcal{C}$.}. Equation~\eqref{Se} agrees with the UV divergent contribution to entanglement entropy in four spacetime dimensions\footnote{In spacetime dimension other than four, the structure of UV divergent contribution to entanglement entropy differs~\cite{Solodukhin:2011}. Hence, extending our approach to arbitrary spacetime dimensions would be nontrivial.}. Notably, this form of the entanglement entropy (albeit not the proportionality constants $1/4l_{\text{P}}^2$ and $\mathcal{C}$) can be derived kinematically, without any assumptions about the gravitational dynamics~\cite{Solodukhin:1995,Mann:1998,Solodukhin:2011}. Moreover, logarithmic corrections to horizon entropy are not limited to the entanglement entropy interpretation. They are predicted by many approaches to quantum gravity, e.g. string theory~\cite{Banerjee:2011}, loop quantum gravity~\cite{Kaul:2000}, AdS/CFT correspondence~\cite{Carlip:2000}, and generalised uncertainty principle phenomenology~\cite{Adler:2001}. Hence, the presence of logarithmic corrections to horizon entropy appears to be a nearly universal feature. For the sake of clarity of presentation, we frame the entire derivation in this paper in terms of the entanglement equilibrium and, therefore, attribute the entropy origin to quantum entanglement. However, our key argument works for any model with logarithmic corrections to entropy, regardless of the entropy interpretation.
	
	To obtain the modified equations for gravitational dynamics, we introduce local causal horizons realised as geodesic local causal diamonds. For simplicity, we consider weak gravitational fields (discarding terms quadratic in the curvature tensors) and assume that all the matter fields are conformally invariant. Then, using the model independent prescription for UV entanglement entropy~\eqref{Se} and the entanglement equilibrium condition, we derive linearised equations for gravitational dynamics.
	
	Let as remark that we already studied the effect of the extra logarithmic term in entropy on the equations for gravitational dynamics in a previous paper~\cite{Alonso:2020b}. In that work, we simplified the derivation by neglecting higher order contributions in the Riemann normal coordinate expansion. Thus, we did not obtain terms with derivatives of the curvature tensors and the correction we found was quadratic in the traceless part of the Ricci tensor. In the present work, we instead include the higher Riemann normal coordinate expansion contributions, but neglect any terms quadratic in the curvature tensor. This can be done consistently, provided that the gravitational field is sufficiently weak. Therefore, we find linearised equations for gravitational dynamics which include correction terms proportional to second derivatives of the curvature tensors. In particular, our result corresponds to linearised equations of motion for quadratic gravity coupled to conformally invariant matter fields, with the higher derivative terms proportional to $l_{\text{P}}^2$. Notably, if the logarithmic correction term in entanglement entropy is negative (as predicted by the majority of approaches), we find that the signs of the higher derivative terms agree with the ones required to avoid tachyonic instabilities~\cite{Salvio:2018,Donoghue:2021}. The emergence of quadratic gravity in this regime can be considered a genuine prediction of the local equilibrium approach, independent of any assumptions concerning the gravitational Lagrangian.
	
	A complete treatment in the strong gravity regime of course requires including both the higher derivative corrections discussed here and the terms quadratic in Ricci tensor, partially derived in our previous paper~\cite{Alonso:2020b}. Moreover, the vacuum case was analysed, suggesting corrections to equations of motion quadratic in the Weyl tensor~\cite{Jacobson:2017}. Presumably, all these terms appear as $O\left(l_{\text{P}}^2\right)$ corrections to the equations for gravitational dynamics. However, such a full treatment is technically very challenging and requires physically well motivated resolution of certain ambiguities (e.g. shape deformations of the horizon). Thus, our current linearised analysis is primarily intended as a proof of concept, showing that local equilibrium conditions provide useful information about gravitational dynamics even beyond the leading order (the Einstein equations). A more complete derivation that includes the terms quadratic in the curvature tensors is forthcoming.

	\section{Derivation of the equations for gravitational dynamics}
	\label{main}
	
	The seminal works concerning the relation of thermodynamics and gravitational dynamics considered local, approximate acceleration horizons (the so called Rindler wedges)~\cite{Jacobson:1995,Chirco:2010,Padmanabhan:2010}. However, these objects are not suitable to perform the computations for the entanglement equilibrium approach, mainly because they do not have a well defined interior to which one can ascribe matter entanglement entropy (and attempts to study entanglement equilibrium of Rindler wedges fail to reproduce the Einstein equations~\cite{Carroll:2016}). Moreover, UV entanglement entropy of Rindler wedges does not have a correction term logarithmic in area (this term is proportional to the Euler characteristic of the surface which is zero for a plane~\cite{Solodukhin:2011}).
	
\begin{figure}[tbp]
		\centering
		\includegraphics[width=.45\textwidth,origin=c,trim={0.1cm 2.4cm 36.7cm 1.5cm},clip]{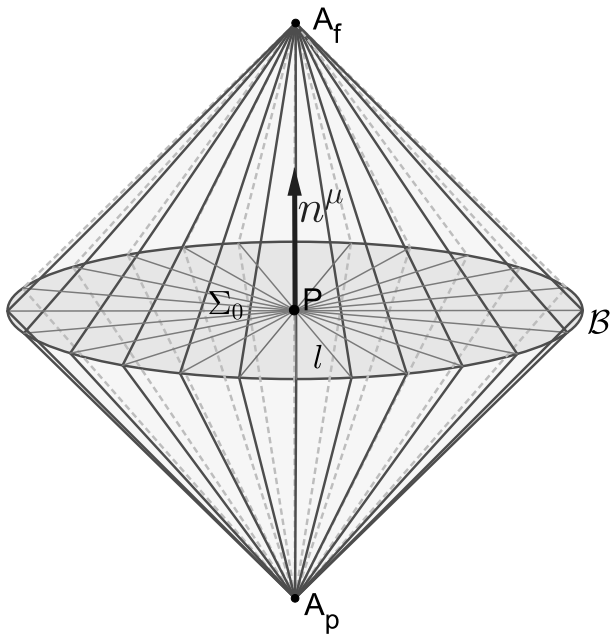}
		\caption{\label{diamond} An illustration of a geodesic local causal diamond centred in $P$. We suppress the angular coordinate $\theta$. The direction of time is given by the unit, future-directed timelike vector $n^{\mu}$. Orthogonal to it is the base of the diamond formed by a 3-dimensional spatial geodesic ball $\Sigma_0$ of radius $l$, whose boundary $\mathcal{B}$ is an approximate 2-sphere. We draw several sample geodesics forming $\Sigma_0$ in grey. The null geodesic generators of the diamond's boundary are represented by the tilted lines starting in the diamond's past apex $A_p$ ($t=-l/c$) and ending in the future apex $A_f$ ($t=l/c$). We can see that $\Sigma_0$ is the intersection of the future domain of dependence of $A_p$ and the past domain of dependence of $A_f$.}
	\end{figure}

	These problems can be solved by instead considering local causal horizons with a closed spatial cross-sections~\cite{Carroll:2016,Svesko:2018,Svesko:2019,Alonso:2020a,Alonso:2020b}. The simplest construction obeying this requirement is a geodesic local causal diamond\footnote{Alternatively, one could use stretched light cones~\cite{Svesko:2018}. We expect the results to be the same in either case.}. A causal diamond centred in an arbitrary regular spacetime point $P$ is fully determined by the choice of a unit, timelike, future-directed vector $n^{\mu}\left(P\right)$ and a length scale $l$. All the geodesics starting from $P$, orthogonal to $n^{\mu}$ and of parameter length $l$ form a 3-dimensional geodesic ball, $\Sigma_0$, provided that $l$ is much smaller than the local curvature length scale (inverse of the square root of the largest eigenvalue of the Riemann tensor). The boundary of $\Sigma_0$, $\mathcal{B}$, is an approximate 2-sphere. The region of spacetime causally determined by  $\Sigma_0$ then forms the geodesic local causal diamond (see figure~\ref{diamond}). To describe it, we use a local Cartesian coordinate system chosen so that $n^{\mu}=\left(\partial/\partial t\right)^{\mu}$.
	
	A causal diamond possesses an approximate (up to curvature-dependent terms) conformal Killing symmetry, generated by the conformal Killing vector field
	\begin{equation}
		\zeta=C\left(l^2-r^2-t^2\right)\partial_{t}-2Crt\partial_{r},
	\end{equation}
	where $r=\sqrt{\sum_{i=1}^3x_i^2}$, and $C$ is an arbitrary real number. On the null boundary of the causal diamond $\zeta^{\mu}$ vanishes and the boundary thus corresponds to a conformal Killing horizon. Moreover, 2-sphere $\mathcal{B}$ (at $t=0$) represents a bifurcation surface as $\zeta^{\mu}$ vanishes there. It has been argued that this horizon has a finite Hawking temperature~\cite{Jacobson:2019}
	\begin{equation}
		\label{TH}
		T_{\text{H}}=\hbar\kappa/2\pi,
	\end{equation}
	where $\kappa=2Cl$ denotes the surface gravity of $\zeta^{\mu}$.
	
	Our starting point is thus a small geodesic local causal diamond defined by a unit, timelike, future-directed vector $n^{\mu}$ and a length parameter $l$. On the one hand, as previously explained, we require $l$ to be significantly smaller than the local curvature scale\footnote{Moreover, to evaluate matter entanglement entropy, $l$ must be smaller than any length scale relevant for the matter fields (e.g. Compton lengths). However, this requirement is not necessary in our case, as we assume the matter fields to be conformally invariant.}. On the other hand, $l$ must be much larger than the Planck length. Otherwise, the quantum gravity effects dominate, making the description of spacetime in terms of a smooth Lorentzian manifold problematic. Within this range, the value of $l$ is arbitrary.

	We require that the diamond is initially in entanglement equilibrium. It has been conjectured that it corresponds to a vacuum, maximally symmetric spacetime, with a local value of the cosmological constant $\lambda$ in principle dependent on the diamond's position and size, i.e., $\lambda=\lambda\left(P,l\right)$~\cite{Jacobson:2016}. In this work, we assume that all the matter fields present in the spacetime are conformally invariant. In that case, it has been shown that, to the leading order, $\lambda$ has a constant value, i.e., $\lambda\left(P,l\right)=\Lambda$~\cite{Jacobson:2016}. Since we introduce $O\left(l_{\text{P}}^2\right)$ corrections to the UV entanglement entropy, the corrections to $\lambda$ will be of the same order,
	\begin{equation}
		\label{lambda}
		\lambda\left(P,l\right)=\Lambda+l_{\text{P}}^2\lambda_{\text{c}}\left(P,l\right).
	\end{equation}

	Upon specifying the equilibrium state of the diamond, we now introduce a small, arbitrary variation of the metric and of the matter fields. Then, the corresponding entanglement entropy variation must vanish to the first order. Let us first discuss the IR, matter fields state-dependent variation of entanglement entropy. For conformal matter fields, the vacuum density matrix can be written in terms of the Hawking temperature~\eqref{TH} and a local modular Hamiltonian, $K$. Then, the IR contribution to the entanglement entropy variation reads~\cite{Jacobson:2016,Bueno:2017,Jacobson:2019}
	\begin{equation}
		\label{dT}
		\delta S_{\text{IR}}=\frac{1}{T_{\text{H}}}\delta K=\frac{2\pi k_Bc}{\hbar}\frac{4\pi}{15}l^4\delta\langle T_{\mu\nu}\rangle n^{\mu}n^{\nu}+O\left(l^5\right),
	\end{equation}
	where $\delta\langle T_{\mu\nu}\rangle$ denotes the expectation value of the energy-momentum tensor variation inside the causal diamond.
	
	Next, we evaluate the UV component of the entanglement entropy variation, which is state independent and fully determined by the variation of the metric. To avoid the technical issues associated with the terms nonlinear  in curvature, we consider only weak gravitational fields with the aim to derive the linearised equations for gravitational dynamics. The weak gravity requirement translates into the Riemann tensor variation $\delta R^{\mu}_{\;\:\nu\rho\sigma}$ being small, so that we can neglect any terms quadratic in the Riemann tensor. More precisely, we demand that any scalar built from two Riemann tensors and the metric is much smaller than the largest scalars obtained from the Riemann tensor, metric tensors and at most two covariant derivatives. We can conveniently describe the metric variation in terms of the Riemann normal coordinate expansion~\cite{Brewin:2009}
	\begin{align}
		\nonumber g_{\mu\nu}\left(x\right)=&\eta_{\mu\nu}-\frac{1}{3}x^{\alpha}x^{\beta}\delta\bar{R}_{\mu\alpha\nu\beta}-\frac{1}{6}x^{\alpha}x^{\beta}x^{\lambda}\nabla_{\rho}\delta\bar{R}_{\mu\alpha\nu\beta} \\
		&-\frac{1}{20}x^{\alpha}x^{\beta}x^{\lambda}x^{\rho}\nabla_{\alpha}\nabla_{\beta}\delta\bar{R}_{\mu\lambda\nu\rho}+O\left(x^5,\delta\bar{R}^2\right). \label{RNC}
	\end{align}
	where
	\begin{equation}
		\delta\bar{R}_{\mu\alpha\nu\beta}=\delta R_{\mu\alpha\nu\beta}+\frac{1}{3}\lambda\left(g_{\alpha\nu}g_{\beta\mu}-g_{\alpha\beta}g_{\mu\nu}\right),
	\end{equation}
	denotes the difference between the Riemann tensor of the maximally symmetric background and the Riemann tensor after the variation of the metric. For the UV contribution to the change of entanglement entropy we straightforwardly obtain from equation~\eqref{Se}
	\begin{equation}
		\delta S_{\text{UV}}=\frac{\delta\mathcal{A}}{4l_{\text{P}}^2}+\mathcal{C}\left(\ln\frac{\mathcal{A}_0+\delta\mathcal{A}}{4l_{\text{P}}^2}-\ln\frac{\mathcal{A}_0}{4l_{\text{P}}^2}\right),
	\end{equation}
	where $\mathcal{A}_0=4\pi l^2$ denotes the flat space area of the 2-sphere $\mathcal{B}$. As we will show in the following, $\delta\mathcal{A}$ is linear in the variation of the Riemann tensor. Hence, we can neglect all the $O\left(\delta\mathcal{A}^2\right)$ terms. The Taylor expansion of the logarithm then yields
	\begin{equation}
		\delta S_{\text{UV}}=\frac{\delta\mathcal{A}}{4l_{\text{P}}^2}+\mathcal{C}\frac{\delta\mathcal{A}}{\mathcal{A}_0}+O\left(\frac{\delta\mathcal{A}^2}{\mathcal{A}_0^2}\right).
	\end{equation}
	The Riemann normal coordinate expansion of the metric allows us to directly obtain the variation of the area element. As we cannot a priori expect that the size parameter $l$ remains constant, the area variation also depends on its variation, $\delta l$. The complete expression for $\delta\mathcal{A}$ reads
	\begin{align}
		\nonumber \delta\mathcal{A}=&\int_{\mathcal{B}}\frac{1}{2}l^{2}\bigg(-\frac{1}{3}x^{\alpha}x^{\beta}\delta\bar{R}_{\mu\alpha\nu\beta}-\frac{1}{6}x^{\alpha}x^{\beta}x^{\rho}\nabla_{\rho}\delta\bar{R}_{\mu\alpha\nu\beta} \\
		&-\frac{1}{20}x^{\alpha}x^{\beta}x^{\lambda}x^{\rho}\nabla_{\alpha}\nabla_{\beta}\delta\bar{R}_{\mu\lambda\nu\rho}\bigg)\text{d}\Omega_{2}+8\pi l\delta l,
	\end{align}
	where on $\mathcal{B}$ we have $x^{\alpha}=lm^{i}\delta^{\alpha}_{i}$, with $m^{i}$ being the outward-pointing radial unit normal to $\partial\Sigma_0$. Using the following identities for spherical integrals
	\begin{align}
		\int_{\mathcal{B}}\text{d}\Omega_{2}m^{i}m^{j}=&\frac{4\pi}{3}\delta^{ij}, \\
		\int_{\mathcal{B}}\text{d}\Omega_{2}m^{i}m^{j}m^{k}=&0, \\
		\int_{\mathcal{B}}\text{d}\Omega_{2}m^{i}m^{j}m^{k}m^{l}=&\frac{4\pi}{15}\left(\delta^{ij}\delta^{kl}+\delta^{ik}\delta^{jl}+\delta^{il}\delta^{jk}\right),
	\end{align}
	we then obtain
	\begin{align}
		\nonumber \delta\mathcal{A}=&-\frac{4\pi l^4}{18}\delta\bar{R}^{ij}_{\;\;\; ij}-\frac{4\pi l^6}{600}\left(\delta^{ij}\delta^{kl}+\delta^{ik}\delta^{jl}+\delta^{il}\delta^{jk}\right) \\
		&\nabla_{i}\nabla_{j}\delta\bar{R}^{m}_{\;\;\, kml}+8\pi l\delta l. \label{area}
	\end{align}

	To proceed, we must fix $\delta l$ which has been so far kept arbitrary. It has been shown that the entanglement equilibrium condition encodes the Einstein equations only if one chooses $\delta l$ so that the volume of $\Sigma_0$ is held fixed~\cite{Jacobson:2016,Jacobson:2019,Jacobson:2019b,Jacobson:2022a,Jacobson:2022b} (since the volume variations enter the first law of causal diamonds). For more general gravitational theories it has been suggested that one should instead fix a local geometric quantity constructed in a precise way from the gravitational Lagrangian and known as the generalised volume, $\mathcal{W}$~\cite{Bueno:2017}. This choice is motivated by the observation that $\mathcal{W}$ plays the role of thermodynamic volume in the first law of causal diamonds. As expected, in general relativity, $\mathcal{W}$ reduces to the geometric volume of $\Sigma_0$.
	
	In our case, since we make no a priori assumptions about the gravitational Lagrangian, we cannot directly derive $\mathcal{W}$. Therefore, for the purposes of the present work, we consider
	\begin{equation}
	\mathcal{W}=\mathcal{V}+l_{\text{P}}^2\mathcal{W}_{\text{q}}\left(P,l,n^{\mu}\right),
	\end{equation}
	where $\mathcal{V}$ is the standard volume of the ball $\Sigma_0$ and $\mathcal{W}_{\text{q}}\left(P,l,n^{\mu}\right)$ an undetermined quantum gravitational correction to it, dependent on the size parameter $l$, the direction of coordinate time $n^{\mu}$ and the spacetime curvature at point $P$. This agrees with the generic form of the generalised volume in modified theories of gravity. We will fix $\mathcal{W}_{\text{q}}$ later on, by requiring independence of gravitational dynamics of our arbitrary choice of time direction $n^{\mu}$. A more refined future treatment should replace the notion of generalised volume by some kinematic, microscopically defined quantity, in the same way  entanglement entropy with a logarithmic correction replaces Wald entropy in the present work. One possibility lies in the proposed relation between the generalised volume and holographic complexity~\cite{Alishahiha:2015,Bueno:2017}.

	We compute the volume variation from the metric expansion~\eqref{RNC} in the same way as the area variation, obtaining for the variation of the generalised volume
	\begin{align}
		\nonumber \delta\mathcal{W}=&\delta\mathcal{V}+\delta\mathcal{W}_{\text{q}}=-\frac{4\pi l^5}{90}\delta\bar{R}^{ij}_{\;\;\; ij}-\frac{4\pi l^7}{4200}\big(\delta^{ij}\delta^{kl}+\delta^{ik}\delta^{jl} \\
		\nonumber &+\delta^{il}\delta^{jk}\big)4\nabla_{i}\nabla_{j}\delta\bar{R}^{m}_{\;\;\, kml}+l_{\text{P}}^2\delta\mathcal{W_{\text{q}}} \\
		&+\left(4\pi l^2+l_{\text{P}}^2\frac{\partial\mathcal{W_{\text{q}}}}{\partial l}\right)\delta l.
	\end{align}
	To satisfy the fixed generalised volume condition, $\delta\mathcal{W}=0$, we impose
	\begin{align}
		\nonumber \delta l=&\frac{1}{1+\frac{l_{\text{P}}^2}{4\pi l^2}\frac{\partial\mathcal{W_{\text{q}}}}{\partial l}}\bigg[\frac{l^3}{90}\delta\bar{R}^{ij}_{\;\;\; ij}+\frac{l^5}{4200}\big(\delta^{ij}\delta^{kl}+\delta^{ik}\delta^{jl} \\
		&+\delta^{il}\delta^{jk}\big)\nabla_{i}\nabla_{j}\delta\bar{R}^{m}_{\;\;\, kml}-l_{\text{P}}^2\delta\mathcal{W_{\text{q}}}\bigg].
	\end{align}
	Since $\mathcal{W}_{\text{q}}$ must vanish in flat spacetime, all the terms containing $\partial\mathcal{W_{\text{q}}}/\partial l$ are already quadratic in curvature. As we only consider linearised equations of motion, we discard them, obtaining
	\begin{align}
	\nonumber \delta l=&\frac{l^3}{90}\delta\bar{R}^{ij}_{\;\;\; ij}+\frac{l^5}{4200}\left(\delta^{ij}\delta^{kl}+\delta^{ik}\delta^{jl}+\delta^{il}\delta^{jk}\right) \\
	&\nabla_{i}\nabla_{j}\delta\bar{R}^{m}_{\;\;\, kml}-\frac{l_{\text{P}}^2}{4\pi l^2}\delta\mathcal{W_{\text{q}}}.
	\end{align}

	Plugging this $\delta l$ into the area variation~\eqref{area}, we have
	\begin{align}
		\label{dA}
		\nonumber \delta\mathcal{A}=&-\frac{4\pi l^4}{30}\delta\bar{R}^{ij}_{\;\;\; ij}-\frac{4\pi l^6}{840}\left(\delta^{ij}\delta^{kl}+\delta^{ik}\delta^{jl}+\delta^{il}\delta^{jk}\right) \\
		&\nabla_{i}\nabla_{j}\delta\bar{R}^{m}_{\;\;\, kml}-\frac{2l_{\text{P}}^2}{l}\delta\mathcal{W_{\text{q}}}.
	\end{align}
	Given that $\lambda$ is constant up to $O\left(l_{\text{P}}^2\right)$ terms (see equation~\eqref{lambda} and the accompanying discussion), we have
	\begin{equation}
		l_{\text{P}}^2\nabla_{i}\nabla_{j}\delta\bar{R}^{m}_{\;\;\, kml}=l_{\text{P}}^2\nabla_{i}\nabla_{j}\delta R^{m}_{\;\;\, kml}+O\left(l_{\text{P}}^4\right).
	\end{equation}
	Since our analysis is insensitive to $O\left(l_{\text{P}}^4\right)$, we discard the derivatives of $\lambda$ in equation~\eqref{dA}, obtaining
	\begin{align}
		\nonumber \delta\mathcal{A}=&-\frac{4\pi l^4}{30}\left(\delta R^{ij}_{\;\;\; ij}-2\lambda\right)-\frac{2l_{\text{P}}^2}{l}\delta\mathcal{W_{\text{q}}} \\
		&-\frac{4\pi l^6}{840}\left(\delta^{ij}\delta^{kl}+\delta^{ik}\delta^{jl}+\delta^{il}\delta^{jk}\right)\nabla_{i}\nabla_{j}\delta R^{m}_{\;\;\, kml}. \label{dA2}
	\end{align}
	The first term on the right hand side can be rewritten in terms of the Einstein tensor, i.e, $\delta R^{ij}_{\;\;\; ij}=2\delta G_{\mu\nu}n^{\mu}n^{\nu}$~\cite{Jacobson:2016}. For the second term, we get, after some straightforward manipulations
	\begin{align}
		\nonumber &\left(\delta^{ij}\delta^{kl}+\delta^{ik}\delta^{jl}+\delta^{il}\delta^{jk}\right)\nabla_{i}\nabla_{j}\delta R^{m}_{\;\;\, kml}=4\nabla_{\mu}\nabla_{\nu}\delta R_{\rho\sigma} \\
		\nonumber &n^{\mu}n^{\nu}n^{\rho}n^{\sigma}+\Big(2\nabla_{\rho}\nabla_{\sigma}\delta R^{\;\:\rho\:\, \sigma}_{\mu\;\:\nu}+4\nabla_{\mu}\nabla_{\rho}\delta R_{\nu}^{\;\:\rho} \\
		\nonumber &+2\nabla^{\rho}\nabla_{\rho}\delta R_{\mu\nu}+\nabla_{\mu}\nabla_{\nu}\delta R\Big)n^{\mu}n^{\nu}+2\nabla_{\rho}\nabla_{\sigma}\delta R^{\rho\sigma} \\
		&+\nabla^{\rho}\nabla_{\rho}\delta R.
	\end{align}

   	Putting together the IR expression~\eqref{dT} and the UV one~\eqref{dA2} for the contributions to the entanglement entropy variation now yields the following entanglement equilibrium condition
	\begin{align}
		\nonumber &\frac{4\pi l^4}{15}\left(\delta G_{\mu\nu}n^{\mu}n^{\nu}-\lambda\right)+\frac{\mathcal{C}l^2l_{\text{P}}^2}{15}\bigg[\delta G_{\mu\nu}n^{\mu}n^{\nu} \\
		\nonumber &+\frac{l^2}{56}\bigg(4\nabla_{\mu}\nabla_{\nu}\delta R_{\rho\sigma}n^{\mu}n^{\nu}n^{\rho}n^{\sigma}+\Big(2\nabla_{\rho}\nabla_{\sigma}\delta R^{\;\:\rho\:\, \sigma}_{\mu\;\:\nu} \\
		\nonumber &+4\nabla_{\mu}\nabla_{\rho}\delta R_{\nu}^{\;\:\rho}+2\nabla^{\rho}\nabla_{\rho}\delta R_{\mu\nu}+\nabla_{\mu}\nabla_{\nu}\delta R\Big)n^{\mu}n^{\nu} \\
		\nonumber &+2\nabla_{\rho}\nabla_{\sigma}\delta R^{\rho\sigma}+\nabla^{\rho}\nabla_{\rho}\delta R\bigg)\bigg]+\frac{2l_{\text{P}}^2}{l}\delta\mathcal{W_{\text{q}}}+O\left(l^5\right) \\
		&=\frac{4\pi l^4}{15}\delta\langle T_{\mu\nu}\rangle n^{\mu}n^{\nu}. \label{balance}
	\end{align}
	The condition contains an arbitrary length scale $l$ (within the previously discussed range) and an arbitrary unit, timelike, future-directed vector $n^{\mu}$. Keeping either of these in the final equations of motion would violate the equivalence principle and the locality of the equations. In the approaches based on Wald entropy, the $l$ and $n^{\mu}$-dependent terms cancel out between the variations of the entropy and the generalised volume~\cite{Bueno:2017}. Ideally, the same procedure should work in our case. However, as discussed above, we do not have a rigorous way to obtain an expression for the generalised volume. Hence, a different approach must be adopted. To deal with the $l$-dependent terms, we construct a sequence of $M_0+1$ causal diamond with sizes $l_{m}=l+m\epsilon l$, where $\epsilon$ is a small dimensionless parameter, $n\in\left[1,M_0\right]$ a natural number ($M_0$ is chosen small enough that $l\left(1+M_0\epsilon\right)$ is much smaller than the local curvature length scale. All the diamonds are centred at the same point $P$ and correspond to the same timelike vector $n^{\mu}$.  In this way, the Riemann normal coordinate  expansion we introduced for the diamond of the size $l$ remains valid. Since the expansion works with values of tensors at $P$, all the tensorial expressions will be the same in each diamond. Only the diamonds size parameter  will differ. To simplify the discussion, we schematically write our equilibrium condition~\eqref{balance} divided into terms of various powers of $l$, i.e.,
	\begin{equation}
	\sum_{m=1}^{\infty}l^{2m}E^{m}=0.
	\end{equation}
	For the $m$-th of the $M_0+1$ causal diamond we constructed, we have the following equilibrium condition 
	\begin{equation}
	\sum_{m=1}^{\infty}\left(1+n\epsilon\right)^{2m}l^{2m}E^{m}=0
	\end{equation}
	We again stress that all $E^{m}$ are the same for each diamond. Then, if the equations are to be satisfied for any $n$, we require $E^{m}=0$ for each $m$\footnote{The correction terms we eventually obtain are proportional to $l_{\text{P}}^2/l_{\text{C}}^2$, with $l_{\text{C}}$ being the local curvature length scale. Then, any terms $l^{2m}E^{m}$ much smaller than this can be neglected, and we only need to set to $0$ finitely many $E^{m}$ using the above described procedure.}. In particular, for $m=2$, we have
	\begin{align}
	\nonumber &\frac{4\pi l^4}{15}\left(\delta G_{\mu\nu}n^{\mu}n^{\nu}-\lambda^{0}\right)+\frac{\mathcal{C}l^4l_{\text{P}}^2}{840}\bigg[4\nabla_{\mu}\nabla_{\nu}\delta R_{\rho\sigma}n^{\mu}n^{\nu}n^{\rho}n^{\sigma} \\
	\nonumber &+\Big(2\nabla_{\rho}\nabla_{\sigma}\delta R^{\;\:\rho\:\, \sigma}_{\mu\;\:\nu}+4\nabla_{\mu}\nabla_{\rho}\delta R_{\nu}^{\;\:\rho}+2\nabla^{\rho}\nabla_{\rho}\delta R_{\mu\nu} \\
	\nonumber &+\nabla_{\mu}\nabla_{\nu}\delta R\Big)n^{\mu}n^{\nu}+2\nabla_{\rho}\nabla_{\sigma}\delta R^{\rho\sigma}+\nabla^{\rho}\nabla_{\rho}\delta R\bigg] \\
	&+\frac{2l_{\text{P}}^2}{l}\delta\mathcal{W_{\text{q}}}^{0}=\frac{4\pi l^4}{15}\delta\langle T_{\mu\nu}\rangle n^{\mu}n^{\nu}. \label{balance 2}
	\end{align}
	where $\lambda^{0}=\lambda^{k}\left(P\right)$ is the piece of the local cosmological constant $\lambda\left(P,l\right)$ independent of $l$ and $\delta\mathcal{W_{\text{q}}}^{0}$ the part of $\delta\mathcal{W_{\text{q}}}$ proportional to $l^5$.
	
	Next, we focus on the dependence on $n^{\mu}$. If the (Einstein) equivalence principle holds, we require the equilibrium condition~\eqref{balance 2} to be independent of $n^{\mu}$. This condition~\eqref{balance 2} contains terms without any contraction with $n^{\mu}$, with two contractions and with four contractions. The terms without any contraction can be rewritten as including two contractions, using the normalisation $g_{\mu\nu}n^{\mu}n^{\nu}=-1$. Regarding the four contraction terms, they must cancel out with the generalised volume variation $\delta\mathcal{W_{\text{q}}}^{k}$. Looking at the generic form of $\delta\mathcal{W_{\text{q}}}^{0}$ in modified theories of gravity, it contains terms of the form $D_{\mu\nu\rho\sigma}n^{\mu}n^{\sigma}h^{\nu\rho}$, where $D_{\mu\nu\rho\sigma}$ is some curvature-dependent tensor and $h^{\nu\rho}=g^{\nu\rho}+n^{\nu}n^{\rho}$ the spatial $3$-metric on $\Sigma_0$. In our case, choosing
	\begin{equation}
	\delta\mathcal{W_{\text{q}}}^{0}=-\frac{\mathcal{C}l^5}{420}\nabla_{(\mu}\nabla_{\nu}\delta R_{\rho\sigma)}n^{\mu}n^{\sigma}h^{\nu\rho}, \label{W}
	\end{equation}
	eliminates the four contraction term, exactly as we require. Sticking to the ansatz $D_{\mu\nu\rho\sigma}n^{\mu}n^{\sigma}h^{\nu\rho}$ for $\delta\mathcal{W_{\text{q}}}^{0}$ this is also the only term we can include, while still maintaining the overall independence of the equilibrium condition~\eqref{balance 2} on $n^{\mu}$.
	
	Plugging expression~\eqref{W} for $\delta\mathcal{W_{\text{q}}}^{k}$ into equation~\eqref{balance} and multiplying by $15/\left(4\pi l^4\right)$ yields an $l$-independent equation which contains two contractions with $n^{\mu}$ (we use that $g_{\mu\nu}n^{\mu}n^{\nu}=-1$)
	\begin{align}
		\nonumber & \delta G_{\mu\nu}n^{\mu}n^{\nu}+\lambda_{\text{rest}}g_{\mu\nu}n^{\mu}n^{\nu}+\frac{\mathcal{C }l_{\text{P}}^2}{56\pi}\bigg(2
		\nabla_{\rho}\nabla_{\sigma}\delta R^{\;\:\rho\;\: \sigma}_{\mu\;\:\nu} \\
		\nonumber &+\frac{4}{3}\nabla^{\rho}\nabla_{\rho}\delta R_{\mu\nu}+\frac{2}{3}\nabla_{\mu}\nabla_{\rho}\delta R_{\nu}^{\;\:\rho}+\frac{2}{3}\nabla_{\nu}\nabla_{\rho}\delta R_{\mu}^{\;\:\rho} \\
		\nonumber &+\frac{1}{3}\nabla_{\mu}\nabla_{\nu}\delta R-2g_{\mu\nu}\nabla_{\rho}\nabla_{\sigma}\delta R^{\rho\sigma}-g_{\mu\nu}\nabla^{\rho}\nabla_{\rho}\delta R\bigg)n^{\mu}n^{\nu} \\
		&=8\pi G\delta\langle T_{\mu\nu}\rangle n^{\mu}n^{\nu}.
	\end{align}
	We now have an equation of the form $f_{\mu\nu}n^{\mu}n^{\nu}$, which holds for any unit, timelike, future-directed vector $n^{\mu}$. It is easy to show that this implies $f_{\mu\nu}=0$ (we include a simple proof in appendix~\ref{proof}). Therefore, we can remove the contractions with $n^{\mu}$
	\begin{align}
		\nonumber & \delta G_{\mu\nu}+\lambda^{k}g_{\mu\nu}+\frac{\mathcal{C }l_{\text{P}}^2}{56\pi}\bigg(2
		\nabla_{\rho}\nabla_{\sigma}\delta R^{\;\:\rho\;\: \sigma}_{\mu\;\:\nu}+\frac{4}{3}\nabla^{\rho}\nabla_{\rho}\delta R_{\mu\nu} \\
		\nonumber &+\frac{2}{3}\nabla_{\mu}\nabla_{\rho}\delta R_{\nu}^{\;\:\rho}+\frac{2}{3}\nabla_{\nu}\nabla_{\rho}\delta R_{\mu}^{\;\:\rho}+\frac{1}{3}\nabla_{\mu}\nabla_{\nu}\delta R \\
		&-2g_{\mu\nu}\nabla_{\rho}\nabla_{\sigma}\delta R^{\rho\sigma}-g_{\mu\nu}\nabla^{\rho}\nabla_{\rho}\delta R\bigg)=8\pi G\delta\langle T_{\mu\nu}\rangle,
	\end{align}
	obtaining $10$ linearised equations for $10$ components of the metric. However, the terms with derivatives of the curvature tensors look fairly complicated, and the equations still contain an undetermined scalar $\lambda^{k}$. We solve the first problem by using the definition of the Weyl tensor
	\begin{align}
		\nonumber \delta C^{\;\:\rho\;\:\sigma}_{\mu\;\:\nu}=&\delta R^{\;\:\rho\;\: \sigma}_{\mu\;\:\nu}+\frac{1}{2}\big(\delta^{\sigma}_{\mu}\delta R_{\nu}^{\;\:\rho}+\delta^{\rho}_{\nu}\delta R_{\mu}^{\;\:\sigma}-g_{\mu\nu}\delta R^{\rho\sigma} \\
		&-g^{\rho\sigma}\delta R_{\mu\nu}\big)+\frac{1}{6}\left(g_{\mu\nu}g^{\rho\sigma}-\delta^{\sigma}_{\mu}\delta^{\rho}_{\nu}\right)\delta R,
	\end{align}
	and the contracted Bianchi identities
	\begin{align}
		\nabla_{\mu}\nabla_{\rho}\delta R_{\nu}^{\;\:\rho}=&\nabla_{\nu}\nabla_{\rho}\delta R_{\mu}^{\;\:\rho}=\frac{1}{2}\nabla_{\mu}\nabla_{\nu}\delta R, \\
		\nabla_{\rho}\nabla_{\sigma}\delta R^{\rho\sigma}=&\frac{1}{2}\nabla^{\rho}\nabla_{\rho}\delta R, \\
		\nonumber \nabla_{\rho}\nabla_{\sigma}C^{\;\:\rho\;\: \sigma}_{\mu\;\:\nu}=&\frac{1}{2}\nabla^{\rho}\nabla_{\rho}R_{\mu\nu}-\frac{1}{6}\nabla_{\mu}\nabla_{\nu}\delta R \\
		&-\frac{1}{12}g_{\mu\nu}\nabla^{\rho}\nabla_{\rho}\delta R.
	\end{align}
	A somewhat lengthy, but straightforward calculation then yields
	\begin{align}
		\nonumber &\delta G_{\mu\nu}+\lambda^{k}g_{\mu\nu}+\frac{\mathcal{C }l_{\text{P}}^2}{56\pi}\bigg(\frac{20}{3}\nabla_{\rho}\nabla_{\sigma}\delta C^{\;\:\rho\;\: \sigma}_{\mu\;\:\nu}+\frac{10}{9}\nabla_{\mu}\nabla_{\nu}\delta R \\
		&-\frac{13}{9}g_{\mu\nu}\nabla^{\rho}\nabla_{\rho}\delta R\bigg)=8\pi G\delta\langle T_{\mu\nu}\rangle. \label{pre eom}
	\end{align}
	To fix $\lambda^{k}$, we assume that energy is locally conserved, i.e., that the energy-momentum tensor is divergenceless, $\nabla_{\nu}T_{\mu}^{\;\:\nu}=0$. Moreover, we have $\nabla^{\nu}\nabla_{\rho}\nabla_{\sigma}\delta C^{\;\:\rho\;\: \sigma}_{\mu\;\:\nu}=0$ up to terms quadratic in the Riemann tensor which we neglect. Then, taking a divergence of equations~\eqref{pre eom} yields
	\begin{equation}
		\nabla_{\mu}\lambda^{k}=\frac{\mathcal{C }l_{\text{P}}^2}{168\pi}\frac{1}{3}\nabla_{\mu}\nabla^{\nu}\nabla_{\nu}\delta R.
	\end{equation}
	Integrating this expression, we find \mbox{$\lambda^{k}=\Lambda+\mathcal{C}l_{\text{P}}^2\nabla^{\rho}\nabla_{\rho}\delta R/\left(168\pi\right)$}, where $\Lambda$ is an arbitrary integration constant corresponding to the cosmological constant. The final result of this derivation are equations for gravitational dynamics corresponding exactly to the linearised equations of quadratic gravity with a cosmological constant, coupled to conformal matter~\cite{Salvio:2018,Donoghue:2021}
	\begin{align}
		\nonumber &\delta G_{\mu\nu}+\Lambda g_{\mu\nu}+\frac{5\mathcal{C }l_{\text{P}}^2}{42\pi}\bigg[\nabla_{\rho}\nabla_{\sigma}\delta C^{\;\:\rho\;\: \sigma}_{\mu\;\:\nu} \\
		&+\frac{1}{6}\left(\nabla_{\mu}\nabla_{\nu}-g_{\mu\nu}\nabla^{\rho}\nabla_{\rho}\right)\delta R\bigg]=8\pi G\delta\langle T_{\mu\nu}\rangle. \label{EoM}
	\end{align}
	The quadratic gravity corrections are proportional to $l_{\text{P}}^2$. This is natural, since $l_{\text{P}}$ is the only constant length scale present in the UV entanglement entropy expression~\eqref{Se}. If $\mathcal{C}<0$, we find that the coupling constants in front of the quadratic gravity terms the signs necessary to avoid the tachyonic instabilities~\cite{Salvio:2018,Donoghue:2021}. Negative $\mathcal{C}$ corresponds to negative logarithmic correction to entanglement entropy of causal diamonds, which is consistent with the results from the replica trick calculations~\cite{Mann:1998,Solodukhin:2011}. The effective action corresponding to equations~\eqref{EoM} reads
	\begin{equation}
	S_{\text{q}}=\frac{1}{16\pi G}\int\left[R+\frac{5\mathcal{C}l_{\text{P}}^2}{168\pi}\left(-\frac{R^2}{3}+C_{\lambda\rho\sigma\tau}C^{\lambda\rho\sigma\tau}\right)\right]\sqrt{-\mathfrak{g}}\text{d}^4x.
	\end{equation}
	We want stress here that the action is recovered only after writing the equations for gravitational dynamics, so it plays no role in their derivation.

	We obtained the quadratic gravity equations of motion from higher order quantum corrections to entanglement entropy. Hence, our result certainly does not support quadratic gravity as the appropriate classical theory of gravity. Instead, the higher derivative terms should be understood as effective field theory corrections to the Einstein equations, which also makes the suppression of the correction terms by $l_{\text{P}}^2$ rather natural. From the effective field theory point of view, the emergence of the linearised equations for quadratic gravity is not surprising. Our derivation assumes the Einstein equivalence principle (this assumption is somewhat implicit here, but discussed at length, e.g. in~\cite{Chirco:2010,Alonso:2020a,Alonso:2020b}), local energy-momentum conservation and purely metric description of gravity (in particular by treating vector $n^{\mu}$ as an arbitrary parameter, rather than a dynamical field). Then, quadratic gravity is the most general theory satisfying these requirements whose equations of motion involve at most fourth derivatives of the metric. The nontrivial insight gained by our derivation is that both the derivatives of the Weyl tensor and the scalar curvature appear in the equations of motion and with the same sign. We stress that here we only analysed the linearised regime, and there is no reason to expect that the full nonlinearised dynamics will be compatible with quadratic gravity.
	
	To conclude our derivation, several technical remarks are in order
	\begin{itemize}
		\item The equations contain the quantum expectation value of the energy-momentum tensor. Therefore, we did obtain linearised equations of motion of semiclassical quadratic gravity. We note that the well-known ambiguity in the definition of the energy-momentum tensor's expectation value has the same form as the quadratic gravity corrections we obtained~\cite{Wald:1994}. Thus, the coefficients in front of these terms in principle depend also on our choice of $\langle T_{\mu\nu}\rangle$.
		\item Our approach cannot be straightforwardly generalised to include nonconformal matter fields. While this is possible in the derivation of Einstein equations, $\lambda$ acquires a potentially large nonconstant contribution due to the effect of nonconformal fields on the IR entanglement entropy~\cite{Jacobson:2016,Casini:2016,Speranza:2016}. Then, the derivatives of $\lambda$  appearing in our construction become non-negligible. Consequently, determining the necessary form of $\lambda$ would become a very complex task. A more refined treatment of both the geometry variations and the IR entanglement entropy is probably necessary to generalise our derivation in this way.
		\item It has been argued in different contexts that logarithmic corrections to horizon entropy also imply modifications to Hawking temperature~\cite{Adler:2001,Solodukhin:2011,Scardigli:2018}. However, we have previously shown that such modifications do not affect the derivation of gravitational dynamics from thermodynamics of local causal horizons~\cite{Alonso:2020b}. This is expected, since both the Unruh and the Hawking effects are purely kinematic~\cite{Visser:2003}, even in the presence of quantum gravitational corrections~\cite{Scardigli:2018}. Therefore, we are justified to use the standard expression for Hawking temperature, regardless of its possible modifications.
		\item Apart from a variation of the size parameter, $\delta l$, one should also in principle allow for variations of the shape of the geodesic ball $\Sigma_0$~\cite{Jacobson:2017}. These deformations contribute to area and volume variations at the order $O\left(l^4\right)$, which indeed affects our analysis. However, their contributions appear to be quadratic in the curvature tensors and, therefore, they do not enter our results~\cite{Jacobson:2017}. This is certainly the case if we allow the shape deformations to depend only on $l$, $n^{\mu}$ the metric, the curvature tensors and their derivatives.
		\item We assumed that the diamond's size $l$ is an arbitrary parameter and the equations of motion are independent of it. Then, the higher derivative terms in the equations of motion appear from the logarithmic term in UV entanglement entropy and are proportional to the Planck length squared ($l_{\text{P}}$ being the only scale available). Alternatively, one might fix $l$ to some $l_{\text{q}}\gg l_{\text{P}}$. Then, the logarithmic correction becomes negligible (it scales with $l_{\text{P}}^2$) and we obtain the higher derivative terms directly from the dominant contribution to entropy, proportional to area. The quadratic gravity corrections then scale with $l_{\text{q}}^2$. This approach is tempting as it foregoes the need for the logarithmic term in entropy and leads to quadratic gravity corrections that become important far below the Planck scale. However, we presently cannot think of any justification for introducing a preferred value of $l$. Therefore, we do not pursue this line of derivation any further.
	\end{itemize}

	\section{Discussion}
	
	In this work, we have derived the linearised equations of motion of quadratic gravity from local equilibrium conditions. In contrast to previous such derivations, we make no a priori assumptions about Wald entropy (and, thus, the gravitational Lagrangian). Instead, our starting point is the Bekenstein entropy with a correction term logarithmic in horizon area, that is nearly universally predicted by both entanglement entropy calculations and by various candidate theories of quantum gravity.
	
	The linearised equations of quadratic gravity are the only equations compatible with the assumptions made in our derivation, i.e., the Einstein equivalence principle and presence of no other fields besides the metric. Hence, the result we arrived at may seem to be trivial. However, it serves as a proof of concept. We have no a priori guarantee that the local equilibrium conditions encode gravitational dynamics beyond the Einstein equations. Our result shows that the leading order linearised corrections can indeed be derived in this way and have the expected form (including the expected signs of the correction term). Therefore, we show that local equilibrium conditions encode nontrivial information about gravitational dynamics beyond the leading order (the Einstein equations), which agree with the expectations one has from the effective field theory approach.
	
	Given this nontrivial emergence of quadratic gravity, it appears to be worthwhile to ask whether local equilibrium conditions can offer any novel insights regarding quantum corrections to gravitational dynamics. In particular, we want to address the questions of diffeomorphism invariance, origin of the cosmological constant, or low energy quantum gravitational effects. Notably, one can ask such questions without committing to an emergent gravity picture, simply treating the local equilibrium as a consistency condition for gravitational dynamics (that is reasonably motivated by the entanglement entropy calculations and standard thermodynamic considerations). For instance, an approach relating thermodynamics of Rindler wedges with loop quantum gravity has been proposed~\cite{Chirco:2014} and one certainly cannot talk about any emergence of gravitational dynamics in this case. Nevertheless, the thermodynamic methods in principle still provide a useful tool to study gravitational dynamics, without the need to use the full machinery of loop quantum gravity. The same is true even if we do not commit to any specific final theory of quantum gravity and simply take the local equilibrium conditions as something any of them should incorporate. Obtaining the linearised equations of motion for quadratic gravity shows that this line of reasoning indeed provides nontrivial insights.
	
	The trade-off for the generality of our derivation lies in the need to fix the generalised volume by an  {\it ad hoc} procedure, requiring that our result is purely metric and respects the equivalence principle. The need to fine-tune the generalised volume in this way presents a drawback that ought to be removed in the future. An improved treatment of the generalised volume and/or shape deformations in our approach might suffice to eliminate the need for fine-tuning the local cosmological constant (as long as the matter fields are conformally invariant).
	
	Let us finally remark that the full, nonlinearised treatment of the entanglement equilibrium is significantly more technically involved. In this case, the shape deformations of the ball $\Sigma_0$ definitely become relevant and a physical principle fixing their contribution is necessary (for a thorough discussion of this issue, see~\cite{Jacobson:2017}). Moreover, the geodesic local causal diamond may not be the appropriate generalisation of the flat spacetime causal diamond for this task~\cite{Wang:2019}. Lastly, it is unclear whether the result will simply be the full quadratic gravity, or a different theory, possibly violating the diffeomorphism invariance. The latter possibility is implied by the previous studies which included the terms quadratic in curvature~\cite{Jacobson:2017,Alonso:2020b}. We will discuss this full treatment in a future work.

	\section*{Acknowledgments}
	
	AA-S is supported by the ERC Advanced Grant No. 740209. ML is supported by the Charles University Grant Agency project No. GAUK 297721. This work is also partially supported by the Spanish Government through Project. No. MICINN PID2020-118159GB-C44.
	
	\appendix
	
	\section{}
	\label{proof}
	
	In this appendix, we explicitly prove that the condition $f_{\mu\nu}n^{\mu}n^{\nu}=0$ for every timelike, unit, future-pointing vector defined in a given point implies $f_{\mu\nu}=0$ (without any loss of generality, we assume $f_{\mu\nu}$ to be symmetric). We introduce a local orthonormal coordinate system such that the metric locally corresponds to the Minkowski one $g_{\mu\nu}=\eta_{\mu\nu}$. We denote the corresponding orthogonal $n$-ad by $n^{\mu}=\left(\partial/\partial t\right)^{\mu}$ and $e_i=\partial/\partial x^i$. Let us stress that, since the Einstein equivalence principle requires that $f_{\mu\nu}$ is a tensor, choosing a specific coordinate system involves no loss of generality. For spacetime dimension $n\ge2$ consider the following set of unit timelike vectors
	\begin{equation}
		t^{\mu}_{ij}=\sqrt{\left(1+p^2+q^2\right)}n^{\mu}+pe_{i}^{\mu}+qe_{j}^{\mu},
	\end{equation}
	for any natural numbers $i$, $j$, such that $0<i<j\le n-1$, and any real numbers $p$, $q$. The requirement $f_{\mu\nu}t^{\mu}_{ij}t^{\nu}_{ij}=0$ for every $t^{\mu}_{ij}$ translates into an equation
	\begin{align}
		\nonumber &\left(1+p^2+q^2\right)f_{00}+p^2f_{ii}+q^2f_{jj}+2p\sqrt{\left(1+p^2+q^2\right)}f_{0i} \\
		&+2q\sqrt{\left(1+p^2+q^2\right)}f_{0j}+2pqf_{ij}=0, \label{f condition}
	\end{align}
	which must be satisfied for any real $p$, $q$. Hence, every coefficient in the expansion of the left hand side in powers of $p$, $q$ needs to be zero. First few conditions implied by this are
	\begin{align}
		f_{00}=&0, \\
		2pf_{0i}=&0, \\
		2qf_{0j}=&0, \\
		p^2\left(f_{00}+f_{ii}\right)=&0, \\
		q^2\left(f_{00}+f_{jj}\right)=&0, \\
		2pqf_{ij}=&0.
	\end{align}
	These equations are satisfied for every $i$, $j$ if and only if $f_{\mu\nu}=0$.

\end{document}